\documentclass[aps,prl,twocolumn,superscriptaddress,showpacs,showkeys]{revtex4}

\usepackage{dcolumn}
\usepackage{amsmath}
\usepackage{graphicx}
\usepackage{rotating}
\usepackage{multirow}

\begin{document}

\newlength{\figurewidth}
\ifdim\columnwidth<10.5cm
  \setlength{\figurewidth}{0.95\columnwidth}
\else
  \setlength{\figurewidth}{10cm}
\fi
\setlength{\parskip}{0pt}
\setlength{\tabcolsep}{6pt}
\setlength{\arraycolsep}{2pt}

\title{Scale-free network growth by ranking}

\author{Santo Fortunato}
\affiliation{School of Informatics, Indiana University, Bloomington, IN 47406, USA}
\affiliation{Fakult\"at f\"ur Physik, Universit\"at Bielefeld, D-33501 Bielefeld, Germany}

\author{Alessandro Flammini}
\affiliation{School of Informatics, Indiana University, Bloomington, IN 47406, USA}

\author{Filippo Menczer}
\affiliation{School of Informatics, Indiana University, Bloomington, IN 47406, USA}

\begin{abstract}
Network growth is currently explained through mechanisms
that rely on node prestige measures, such as degree or fitness.
In many real networks 
those who create and connect nodes do not know the prestige values of 
existing nodes, but only their ranking by prestige.
We propose a criterion of network growth that explicitly relies on the
ranking of the nodes according to any prestige measure, be it topological 
or not.
The resulting network has a scale-free degree distribution when 
the probability to link a target node is any power law function of its rank,  
even when one has only partial information of node ranks. 
Our criterion may explain the frequency and robustness of scale-free
degree distributions in real networks, as illustrated by the special 
case of the Web graph.
\end{abstract}



\maketitle

Scholars have become interested in the many complex networks 
with long-tailed degree distribution~\cite{DM02,bara,newman} 
due to their peculiar structural features like 
resilience~\cite{bara1} and to the critical dynamical processes 
taking place on these networks, including epidemics
spreading~\cite{PV01a}, search~\cite{ALPH01}, and opinion
formation~\cite{extreme}.
The most popular explanation for the origin of scale-free networks
is \emph{preferential attachment}~\cite{aba},  
according to which 
a newly created node is connected to a
pre-existing one with a probability \emph{exactly proportional} to the
number of links (degree) of the target node.  This mechanism embodies
the intuitive idea of a `rich get richer' dynamics.  In the
limit of infinitely many nodes, the degree distribution 
of the resulting network has a power law tail with exponent $\gamma=3$.
Preferential attachment has been explicitly or implicitly embodied in
many successive models of network growth~\cite{copy,doromod}. 

Krapivsky and Redner~\cite{redner} showed that the proportionality 
between linking probability and target node degree is
a necessary ingredient of preferential attachment; 
if the linking probability is 
a power of the degree with exponent $\alpha$, the resulting network
has a power law degree distribution only when $\alpha=1$.  For
$\alpha<1$ the degree distribution is a power law multiplied by a
stretched exponential and for larger values of $\alpha$ the model yields
star-like networks.
This 
seems at odds with the abundance and robustness of scale-free degree
distributions in real networks.  
Other proposed mechanisms do not rely on preferential attachment. 
If the
attraction of links depends on some ``fitness'' property of 
the target node, the networks
display scale-free degree distributions for some suitable
choices of the fitness distribution~\cite{calda,marian03}.  

Whether the link attractiveness of a node depends on a 
prestige measure exogenous or endogenous to the network topology, 
this information may not be available in real cases. 
In a social network, for example, we could assume that the
probability for a person to make new friends is proportional to
properties such as popularity, 
attractiveness, 
or wealth --- all typically difficult for strangers to
quantify, measure, or discover.

While the \emph{absolute} importance of an object is often unknown, 
it is quite common to have a clear idea about the \emph{relative}
values of two objects.  One can often say who is the richer or more
popular between two individuals. 
As another example, search engine users only see how Web pages are
ranked.  The perception of how items are ranked 
requires far less information than their actual importance.
Here we propose a model of network growth that focuses on the relative
rather than absolute importance of the nodes, which  
are ranked according to an arbitrary prestige measure. 
We show that 
scale-free networks emerge for a very general form 
of the linking probability
and are stable for a large 
range of the parameters describing the growth. 
The result holds even if 
new nodes have information on 
only subsets of older nodes. 

First, a prestige ranking criterion is selected.
At the $(t+1)$-th iteration, the new node $t+1$ is created and new
links are set from it to $m$ pre-existing nodes.  The previous
$t$ nodes are ranked according to prestige, and the linking
probability $p({t+1}\rightarrow j)$ that node $t+1$ be connected to
node $j$ only depends on the rank $R_j$ of $j$: 
\begin{equation}
p({t+1}\rightarrow j) =\frac{{R_j}^{-\alpha}}{\sum_{k=1}^t {R_k}^{-\alpha}}
\label{eq1}
\end{equation}
where $\alpha>0$ is a real-valued parameter.
The linking probability clearly decreases with increasing rank.

The choice of prestige measure is arbitrary.  
We discuss both topological measures (age $t$ and degree $k$) 
and exogenous ones (any node fitness $\eta$).

If the nodes are sorted by age, from the oldest to the newest, 
the label of each node coincides with its rank, 
i.e.
$R_t=t$ $\forall t$. In this special case, our linking probability 
coincides with that of the so-called static network model~\cite{kahng}.
We can calculate the number of
links that the $R$-th node will attract since its creation.  
Suppose that the evolution of the network stops when $N$ nodes are
created.  At each iteration a constant number $m$ of links are created
between the new node and the older ones.  The expected total number
$k^N_R$ of links that the $R$-th node has attracted at the end is
\begin{equation}
k^N_R=\sum_{t=R+1}^N \frac{m{R}^{-\alpha}}{\sum_{j=1}^t {j}^{-\alpha}}.
\label{eq2}
\end{equation}
The first sum runs over $N-R$ terms because $N-R$ nodes are created
after node $R$, and each of them can be connected to $R$.  We now
approximate the sums with integrals and assume that $N \gg R$, as we are
ultimately interested in the thermodynamic limit.  
We find that $k^N_R = A R^{-\alpha}$, where $A$ is a
function of $N$, $\alpha$ and $m$. 
Knowing $k^N_R$ it is possible to 
find how many nodes $N_{k}$ have
the same expected number of links $k$.  The ratio $N_{k}/N$
in the limit of large $N$ yields the  
probability $p(k,N)$ that a node of the network has degree $k$:
\begin{equation}
p(k,N) \sim k^{-(1+1/\alpha)}.
\label{eq6}
\end{equation}
Eq. \ref{eq6} shows that the
degree distribution of the network follows a power law with exponent
\begin{equation}
\gamma=1+1/\alpha 
\label{eq7}
\end{equation}
for any value of $\alpha$.  Since $\gamma$ can take any value
greater than $1$, we can in principle reproduce the exponents
measured in real systems.  
For $\alpha>1$ ($\gamma < 2$) a few nodes attract a finite 
fraction of all links (condensation); 
in the limit case in which the power law of Eq.~\ref{eq1}
is replaced by a simple exponential  
the network still has a long-tailed degree distribution 
with $\gamma=1$ (as in the limit $\alpha\rightarrow\infty$).  

%

\begin{figure}[t]
    \includegraphics[width=\figurewidth]{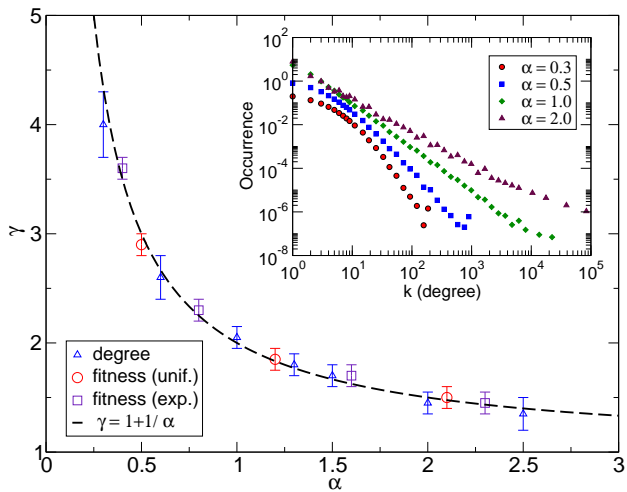}%
    \caption{\label{fig1} Inset: In-degree distributions for networks built according to 
    our model (ranking by degree). 
    The number of nodes is $N=10^6$. 
    Data are averaged within logarithmic bins of degree
	and shifted along the $y$-axis for
	illustration purposes. The main curve plots the degree 
	distribution exponent $\gamma$ 
	as a function of $\alpha$, from simulations with nodes ranked by 
	various criteria. Error bars represent standard errors 
	on the best fit estimates of $\gamma$, while the dashed 
	line is the prediction of Eq.~\ref{eq7}.}
\end{figure}

Let us now consider a more realistic ranking criterion, 
the in-degree. The number  
of incoming links of a node represents how many
times the node has been selected by its peers.  
For undirected networks we can equivalently use the degree. 
Nodes with 
(in-)degree zero, which if present are a problem for the extension of
other growth models to directed networks, do not raise an issue here
because they have ranks expressed by positive numbers, like
all other nodes.

%
To see what kind of networks emerge with this new prestige measure,
we cannot apply the above derivation 
because the degree-based ranking of a node can change over time.
On the other hand, for a growing network there is a strong correlation
between the age of a node and its degree, as older nodes have more
chances to receive links.  Furthermore the ranking of nodes according
to degree is quite stable~\cite{lead}. 
Therefore we expect the same result as for the ranking by age.

To verify our expectation we performed Monte Carlo simulations of 
the network growth process with the new degree-based ranking strategy. 
The inset of Fig.~\ref{fig1} shows the degree distributions of four
networks, corresponding to various values of the exponent $\alpha$.
In the 
logarithmic scale of the plot the tails appear as straight lines, as
one would expect for scale-free distributions.
To verify that the relation between $\alpha$ and the exponent
$\gamma$ of the degree distribution is the one predicted by our model, 
in the main plot of Fig.~\ref{fig1} we compare  
various pairs ($\alpha$, $\gamma$) with the hyperbola of
Eq.~\ref{eq7}.  The agreement is evident.

A striking feature of our model is that it generates 
scale-free networks even when ranking nodes by a prestige measure 
unrelated to the network topology, i.e. by some exogenous fitness 
attribute of the nodes. Suppose we assign a fitness
$\eta$ to the nodes according to an arbitrary distribution. 
Let $R_j(t)$ be the expected rank of node $j$, with fitness $\eta_j$,
among $t$ nodes. As $R_j(t)$ is asymptotically proportional to $t$, 
we can write $R_j(t)=t\,\rho(\eta_j)$, where the relative rank 
$\rho(\eta_j)$ depends only on $\eta_j$. 
By replacing the rank in the linking probability of 
Eq.~\ref{eq1} by the expression $t\,\rho(\eta_j)$, we can factor out 
$t^{-\alpha}$ from both the numerator and the denominator, 
leading us back to the case of static ranking discussed above.

Monte Carlo simulations confirm the result.
We used uniform, exponential and power law fitness distributions. 
The resulting networks have degree distributions with
power law tails, for any value of $\alpha$ and any fitness distribution.
The relation between the exponents $\gamma$ and $\alpha$ is 
in agreement with the prediction of Eq.~\ref{eq7} in all 
cases. In Fig.~\ref{fig1} we 
illustrate this relation for the uniform and exponential cases.

Most models of network growth assume that a new node can be linked to
any existing node, chosen according to some criterion.  This requires
that the new node be aware of the status of all its peers.  Such an
assumption of complete knowledge of the network may not be realistic.
For instance, in a large social network
nobody knows everybody else.  It is reasonable to suppose that
the knowledge that each node has of the network is limited to a sample
of nodes, which is in general different from node to node.  In the
previous example, each person has his/her own group of acquaintances.
Having access only to portions of a network can profoundly affect the
dynamics of network growth.  Mossa \textit{et al.}~\cite{mossa} showed
that preferential attachment would yield power law degree
distributions ending with exponential cutoffs.

Let us check whether and to which extent the hypothesis of limited
information affects the dynamics of the rank-driven model proposed
here.  We assume that whenever a new node is created, it can `see' each
of the preexisting nodes with a probability $h$.  We shall see that if
$h$ is constant, our earlier
result 
holds; if $h$ is power-law distributed, the scenario is more complex, 
but one recovers long-tailed degree distributions in most cases. 

If $h$ is constant the size distribution of the subsets accessed by
the new nodes is a binomial peaked at $ht$, where $t$ is the
number of nodes of the network after $t$ iterations.  This means that
most subsets will have a size of about $ht$.  Let us assume that the nodes
are ranked according to
their age,  
making a formal analysis possible.  The linking
probability is still given by Eq.~\ref{eq1}, with the important caveat
that now we deal only with the nodes within the subset accessed by the
newly created node.  So the ranks of Eq.~\ref{eq1} refer to the
ordering of the nodes of the subset, not of all nodes like
before.  Let us indicate the `local' ranks with $r$, to distinguish them
from the global ranks $R$ we have dealt with so far.

When a new node $t \! + \! 1$ is created, it knows a list of $n$ older nodes. 
One can calculate the probability 
$p(R,r,t,n,h)$ 
that node $R$ has local
rank $r$ within this list,   
and from this 
the probability 
$p({t \! + \! 1}\rightarrow R,r,n,h)$ 
that such a node be
linked to $t \! + \! 1$. 
Then we 
sum over the possible ranks $r$ of node
$R$ in the list ($r \in 1 \dots n$) and all possible subset sizes
($n \in 1 \dots t$).  The result yields the linking probability
$p({t+1}\rightarrow R,h)$ of $t+1$ to $R$:
\begin{eqnarray}
\lefteqn{p({t \! + \! 1}\rightarrow R,h)=}\nonumber \\
& &\sum_{n=1}^{t}\sum_{r=1}^{n}\frac{h^n(1-h)^{t-n}r^{-\alpha}}{\sum_{m=1}^{n} m^{-\alpha}} 
\left(\begin{array}{c}R-1 \\
r-1 \\
\end{array} \right)
\left(\begin{array}{c}t-R \\
n-r \\
\end{array} \right).
\label{eq20}
\end{eqnarray}
From Eq.~\ref{eq20} we see that if $h=1$, which corresponds to a
list with all $t$ nodes, one recovers Eq.~\ref{eq1} as expected.  For
$h<1$, however, it is not possible to derive a close expression for
$p({t+1}\rightarrow R,h)$, so 
we performed Monte Carlo simulations of the process leading to
Eq.~\ref{eq20}. 
In every simulation we produced a large number of lists, each 
formed by sampling nodes with probability $h$. 
At the beginning of the simulation we initialized all entries of the
array $p({t+1}\rightarrow R,h)=0$.  Once a list was completed, we added
to the entries of $p({t+1}\rightarrow R,h)$, corresponding to the nodes
of the list, the linking probability as given by Eq.~\ref{eq1} (with
the proper normalization).  With this method we simulated systems with
up to $N=10^6$ nodes.

For $t$ not too small, $p({t+1}\rightarrow R,h)$ is well approximated by the 
following function:
\begin{equation} 
p({t \! + \! 1}\rightarrow R,h) \sim \left\{ 
  \begin{array}{ll}
    \frac{(\alpha-1)h^\alpha}{\alpha\,h^{\alpha-1}-t^{1-\alpha}} & \mbox{if $1 \leq R \leq [\frac{1}{h}]$}\\
    \frac{(\alpha-1)R^{-\alpha}}{\alpha\,h^{\alpha-1}-t^{1-\alpha}} & \mbox{if $[\frac{1}{h}] \leq R \leq t $}.
  \end{array} \right.
\label{eq22}
\end{equation}
The first $[\frac{1}{h}]$ nodes have the same probability of being
selected. For the other nodes, $p({t+1}\rightarrow R,h)$ has
the same dependence on $R$ as in the case in which there is complete
information on the network ($h=1$).  This means that in a network grown
with the linking probability of Eq.~\ref{eq20}, the first
$[\frac{1}{h}]$ nodes will have approximately the same number of links,
whereas the degrees of the others will have the distribution of
Eq.~\ref{eq6}.  If $h$ is independent of $N$, the subset of equiprobable 
nodes does not grow with $N$ and has no structural 
relevance when $N \rightarrow \infty$; 
if $h \sim 1/N$ the subset of equiprobable nodes is a fraction of 
the network and the degree distribution has an exponential cutoff. 
Monte Carlo simulations confirmed the result.  We conclude that the
degree distribution of networks grown with our ranking strategy is the
same whether new nodes have access to the full network or just to
subsets of it, as long as the subsets contain a constant proportion of the 
network nodes.  The
latter assumption may not be realistic, as the number of contacts may
vary appreciably from node to node.  Next we extend our analysis to
this case.

In general, if $h$ is distributed according to a function $S(h)$, we
need to convolute the $p({t+1}\rightarrow R,h)$ of Eq.~\ref{eq20} with
$S(h)$ to get the linking probability $p_S({t+1}\rightarrow R)$ of the
full process,
\begin{equation} 
p_S({t+1}\rightarrow R)=\int_{h_m}^{h_M}S(h)p({t+1}\rightarrow R,h)dh,
\label{eq23}
\end{equation}
where $h_m$ and $h_M$ are the extremes of the interval where $S(h)$ is
defined.  We consider the following simple probability distribution for
$h$:
\begin{equation}
S(h)=\frac{\beta-1}{t^{\beta-1}-1}h^{-\beta}
\label{eq24}
\end{equation}
where $\beta>0$.  The function is defined in the range $h \in [1/t,
1]$.  In fact, for a network with $t$ nodes, in order to have at least
one item in a random selection of nodes one needs a probability $h \geq
1/t$.  The prefactor ensures the normalization of the function in the
interval $[1/t, 1]$.  As we discuss later, the choice of the power law
in Eq.~\ref{eq24} is a realistic one; it also accounts for the two
limit cases of uniform ($\beta=0$) and exponential ($\beta
\rightarrow \infty$) distributions; finally it allows us to treat the
problem analytically, provided we introduce reasonable approximations.

We plug 
Eqs.~\ref{eq22} and \ref{eq24} into Eq.~\ref{eq23}. 
With $R$ fixed, we split the integral over the two $h$ domains corresponding 
to the regimes of Eq.~\ref{eq22}.
Depending on the values of the parameters $\alpha$ and $\beta$, 
we can neglect different terms in the resulting integrands, leading to  
four cases for the asymptotic dependence of the linking 
probability on rank.

\begin{figure}[t]
    \centerline{\includegraphics[width=0.92\figurewidth]{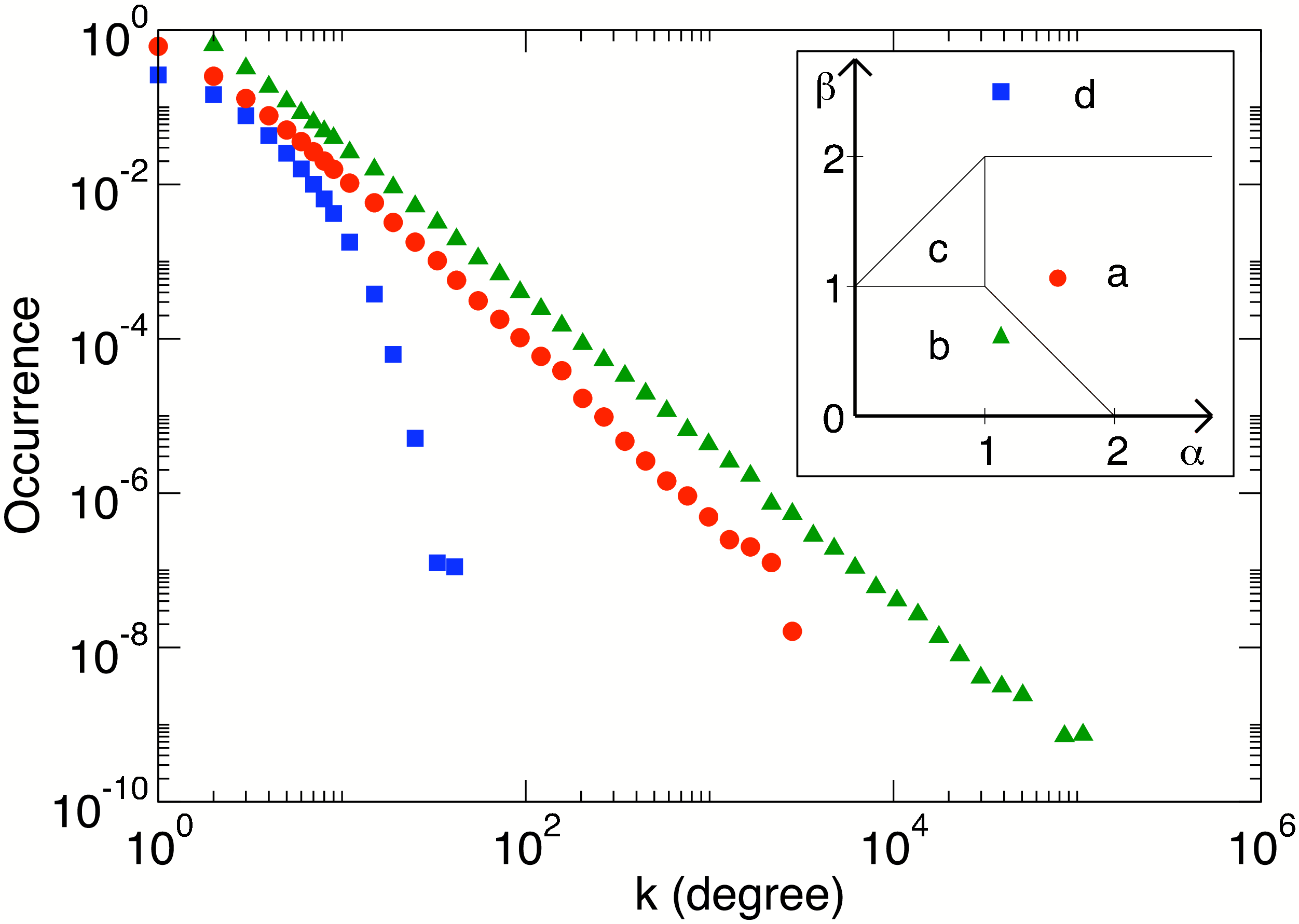}}
    \vspace{-10pt}
    \caption{\label{map} Degree distributions for networks 
    grown according to the proposed model, with incomplete information. 
    Inset: Regions of the parameter space for the four 
    regime cases. The marked points correspond to the ($\alpha, \beta$) 
    values that generate the distributions in the main plot. 
    }
\end{figure}

In three of the cases 
$p_S({t+1}\rightarrow R)$ has a power law dependence on $R$: the
networks grown for the corresponding values of $\alpha$ and $\beta$
will then have scale-free degree distributions.
The power law distribution of the probability $h$ affects the value of
the exponent $\gamma$ of the degree distribution, which no longer
depends only on $\alpha$ 
and in one case 
is only a function of
$\beta$.  In the last case the linking probability is independent
of the rank, so all nodes have the same chance of being linked and
the degree distribution is exponential.  
The four cases correspond to regions of the 
($\alpha, \beta$) parameter quadrant (Fig.~\ref{map} inset). 
In cases $a$, $b$ and $c$, 
$p(k) \sim k^{-\gamma}$ with $\gamma_{a} = 1+1/(2-\beta)$, 
$\gamma_{b} = 1+1/\alpha$, and $\gamma_{c} = 1+1/(1+\alpha-\beta)$. 
In case $d$, $p(k)$ is exponential. 
Monte Carlo simulations 
confirm 
these predictions, as shown in Fig.~\ref{map}. 
The results are identical if
the nodes are ranked according to degree or fitness.

Compared to mechanisms proposed in the past to explain the 
emergence of scale-free networks, the rank-based model introduced 
here presents three main advantages. First, it assumes less information 
is available to nodes (or node creators); it seems more realistic 
in many real cases to imagine that the relative importance of 
items is easier to access than their absolute importance. 
Second, the link attractiveness of nodes is by no means restricted to 
topology; it can depend on exogenous attributes of the nodes, 
which makes our model suitable for applications in many different contexts.
Third, the criterion is more robust in that: (i) it naturally 
extends to directed networks; (ii) it leads to long-tailed 
degree distributions for a broad class of linking probability 
functions --- namely power laws of rank with any exponent 
$\alpha > 0$, including the degenerate exponential case for 
$\alpha \rightarrow \infty$; and (iii) the scale-free degree 
distribution generated by the model is not affected by limiting 
the information available to subsets of nodes. 

The rank-based model is directly applicable to the Web as a special
case, if one considers the role of search engines in the discovery of
pages~\cite{Fortunato05egalitarian}.  When a user submits a query, the
search engine ranks the results by various criteria including a
topological prestige measure, PageRank, 
closely correlated with in-degree.  Users do not know the PageRank of the
search hits, but observe their ranking and thus are more likely to
discover and link pages that are ranked near the top.  
By analyzing search engine logs one finds that (i) 
users click on result hits with a probability that is a power law 
function of rank matching Eq.~\ref{eq1}, with $\alpha = 1.6$;  and (ii) user
queries return hit sets whose size distribution matches the power law
in Eq.~\ref{eq24}, with $\beta = 1.1$.  
Assuming that users tend to
link pages that they discover by searching, and that they are only
aware of the pages returned by search engines in response to their
queries, our model predicts a scale-free distribution of in-degree
with exponent $\gamma_{a} = 2.1$ (cf. case $a$ and curve marked 
with circles in Fig.~\ref{map}). This is in
perfect agreement with established Web measurements~\cite{bara1}.

\begin{acknowledgments}
We thank 
A. Vespignani, M. Serrano, M.
Bogu\~{n}\'{a}, V. Colizza and M. Barth\'elemy for helpful discussions.  Work funded in part
by a Volkswagen Foundation grant to SF, NSF award 0348940 to
FM, and Indiana University School of Informatics.
\end{acknowledgments}

\end{document}